\documentclass{ifac/ifacconf}

\usepackage{graphicx}      % include this line if your document contains figures
\usepackage{natbib}        % required for bibliography
\usepackage{comment}
\usepackage{amsfonts}
\usepackage{amsmath}
\usepackage{subcaption}
\usepackage{lipsum}
\usepackage{tikz}
\usepackage{float}
\usepackage{url}
\usetikzlibrary{positioning, shapes.geometric, arrows.meta}

\begin{document}
\begin{frontmatter}

\title{A Fairness-Oriented Multi-Objective Reinforcement Learning approach for Autonomous Intersection Management\thanksref{footnoteinfo}} 

\thanks[footnoteinfo]{This work was partially carried out within the Italian National Center for Sustainable Mobility (MOST) and received funding from NextGenerationEU (Italian NRRP – CN00000023 - D.D. 1033 17/06/2022 - CUP C93C22002750006).}

\author{Matteo Cederle,} 
\author{Marco Fabris,} 
\author{Gian Antonio Susto}

\address{University of Padova, 
   Italy (e-mail: matteo.cederle@phd.unipd.it, marco.fabris.1@unipd.it, gianantonio.susto@unipd.it).}
%\address[Second]{Colorado State University, 
%   Fort Collins, CO 80523 USA (e-mail: author@lamar. colostate.edu)}
%\address[Third]{Electrical Engineering Department, 
%   Seoul National University, Seoul, Korea, (e-mail: author@snu.ac.kr)}

\begin{abstract}                % Abstract of 50--100 words

This study introduces a novel multi-objective reinforcement learning (MORL) approach for autonomous intersection management, aiming to balance traffic efficiency and environmental sustainability across electric and internal combustion vehicles. The proposed method utilizes MORL to identify Pareto-optimal policies, with a post-hoc fairness criterion guiding the selection of the final policy. Simulation results in a complex intersection scenario demonstrate the approach's effectiveness in optimizing traffic efficiency and emissions reduction while ensuring fairness across vehicle categories. We believe that this criterion can lay the foundation for ensuring equitable service, while fostering safe, efficient, and sustainable practices in smart urban mobility.
\end{abstract}


\end{frontmatter}
%===============================================================================

\section{Introduction}
\label{sec:intro}

%{\color{red}\lipsum[1-4]}

Autonomous intersection management (AIM) is gaining increasing attention as a promising paradigm to improve traffic throughput and safety in urban mobility systems (\cite{dresner2008multiagent,liu2017distributed,klimke2022enhanced}). Recent advances in Reinforcement Learning (RL) have enabled the development of flexible and adaptive controllers capable of managing the complex and stochastic dynamics of intersection scenarios (\cite{vidali2019deep,klimke2023automatic,Cederle2024ATT}). However, most existing AIM approaches prioritize global performance objectives—such as minimizing travel time—without accounting for broader societal concerns, such as environmental sustainability and fairness among heterogeneous road users.

The latest research has demonstrated that RL can be used not only to optimize operational efficiency in smart mobility systems, but also to promote fairness in service delivery. In the context of shared micromobility systems, for example, fairness-aware RL frameworks have been proposed to reduce service disparities between high- and low-demand areas by incorporating spatial equity metrics into the reward function (\cite{cederle2025regulatingspatialfairnesstripartite,cederle2025fairnessorientedreinforcementlearningapproach}). These methods highlight the potential of learning-based control to embed fairness objectives directly into system operation.

Translating this novel perspective to AIM, fairness pertains to the temporal and categorical distribution of access to the intersection. Without explicit fairness constraints, RL agents trained to optimize throughput or environmental impact may systematically favor certain classes of vehicles—such as those with faster dynamics or shorter paths—resulting in persistent delays for others. This is especially relevant in heterogeneous traffic scenarios involving electric and internal combustion vehicles, or private and public transport (\cite{fleming2018social,yan2020fairness}). To address this, recent studies have begun to explore fairness-aware RL mechanisms that adjust priority or scheduling decisions based on service equity considerations across agents (\cite{li2020fairness,wu2023traffic}). These approaches suggest that integrating fairness into AIM policies is %not only feasible, but 
necessary to ensure socially responsible automation in urban traffic management.

In parallel, the field of multi-objective reinforcement learning (MORL) has provided tools to capture trade-offs between competing objectives (\cite{roijers2013survey,felten2024multi}). Despite its theoretical maturity, MORL has seen limited adoption in safety-critical mobility applications, where the explicit handling of multiple objectives—such as traffic efficiency and emissions reduction—can be crucial to support ethical and sustainable deployment.
This work addresses the above gap within the challenge of fairness in AIM, where optimizing for efficiency or environmental goals alone may lead to systematically biased service across vehicle categories (e.g., petrol vs. electric). Specifically, we propose a multi-objective learning-based framework that enables equitable decision-making by incorporating a post-hoc criterion to assess fairness across Pareto optimal solutions. Such an approach promotes a more balanced distribution of traffic flows while preserving safety, efficiency, and sustainability, thus reducing the need for highly-interconnected monitoring infrastructure (see, e.g., the analysis in \cite{fabris2025efficient}). 

%We believe that this promotes a more balanced distribution of traffic flows while preserving efficiency and sustainability, thus reducing the need for highly-interconnected monitoring infrastructure (see, e.g., the analysis in \cite{fabris2025efficient}). 

The main contributions brought by this paper can be summarized as follows:
\begin{enumerate}
	\item We propose the first MORL based approach for autonomous intersection management.
	\item We test our method in a simulated environment and we show its effectiveness in learning a set of safe Pareto optimal policies, balancing the trade-off between efficiency and environmental sustainability.
	\item We introduce a \textit{post-hoc} fairness of service criterion, useful to guide the selection of the final policy to be deployed in the environment.
\end{enumerate}
The remainder of this manuscript unfolds as follows: Section \ref{sec:backgr} reviews some key concepts in deep RL and the graph-based representation of the AIM scenario introduced in previous works. Section \ref{sec:method} presents the core of our contribution, while Section \ref{sec:exp} shows the results achieved through several virtual experiments. Finally, Section \ref{sec:conc} concludes our work and presents some directions for future improvement.

\section{Background}
\label{sec:backgr}

\subsection{Deep Reinforcement Learning}

RL is a paradigm in which an \textit{agent} acquires the ability to solve a \textit{task} through repeated interactions with its \textit{environment}. The objective is to maximize the \textit{cumulative rewards} obtained over time. At the core of RL lies the framework of the \textit{Markov Decision Process} (MDP, \cite{sutton2018reinforcement}), a formalism that defines the problem as a five-element tuple: ${\left\langle \mathcal{S}, \mathcal{A}, \mathcal{P}, \mathcal{R}, \gamma \right\rangle}$. Here, $\mathcal{S}$ and $\mathcal{A}$  represent the \textit{state} and \textit{action spaces}, respectively. The agent interacts with the environment through a policy $\pi:~\mathcal{S} \to \mathcal{A}$, selecting actions that influence future states. The \textit{state transition function} $\mathcal{P}$ models the system’s dynamics: 
$\mathcal{P}(s, a, s') = P\left[S_{t+1} = s' \mid S_t = s, A_t = a \right]$.
Each transition is evaluated through a \textit{reward function} $\mathcal{R}(s, a, s'): \mathcal{S} \times \mathcal{A} \times \mathcal{S} \to \mathbb{R}$, which provides feedback on the quality of the chosen action. Additionally, the discount factor $\gamma \in [0,1)$ regulates the importance of future rewards, shaping the cumulative expected return under policy $\pi$, i.e.
$J(\pi)=~\mathbb{E}_\pi[\sum_{t=0}^\infty \gamma^tr_{t+1}]$.
Solving an RL problem coincides with finding the optimal policy that maximizes $J(\pi)$. In high-dimensional or complex environments, the policy is often parameterized by a neural network with parameters $\phi$, i.e. $\pi\to\pi_\phi$, optimized via gradient ascent exploiting the well-known Policy Gradient Theorem:
\begin{equation}
	\nabla_\phi J(\pi_\phi) = \mathbb{E}_\pi \left[ \nabla_\phi \log \pi_\phi(a|s) Q^{\pi_\phi}(s,a) \right],
\end{equation}
\begin{equation}
	\phi \leftarrow \phi + \alpha \nabla_\phi J(\pi_\phi),
\end{equation}
where $\alpha$ is the learning rate. In practice, also the action-value function $Q^{\pi_\phi}(s,a)$ is approximated by a neural network with parameters $\theta$, i.e. $Q\to Q_\theta$, leading to the widely used family of actor-critic algorithms (\cite{konda1999actor}).

In this work, vehicle control is achieved by regulating longitudinal acceleration within the range $[a_{\min},a_{\max}]$, requiring an RL algorithm tailored for continuous action spaces. Similar to previous works, we employ the state-of-the-art actor-critic algorithm Twin Delayed Deep Deterministic Policy Gradient (TD3, \cite{fujimoto2018addressing}). We refer to the original paper for a detailed description of its implementation.

\subsection{Graph-based Autonomous Intersection Management}
\label{subsec:graphaim}

In this section we will briefly review the original graph-based representation of the AIM scheme, proposed in \cite{klimke2022enhanced, klimke2022cooperative}. The state of the environment at each time step is encoded by a graph, denoted by $(\mathcal{V,E,U})\in\mathcal{S}$. Each vehicle in the scene is mapped to a vertex $\nu$ in the set $\mathcal{V}$, each of which stores the corresponding input features $h_\nu=[s,v,\tilde{a}]^\top$, representing the longitudinal position, scalar velocity and measured acceleration\footnote{The tilde symbol is employed to distinguish between the measured acceleration $\tilde{a}$ and the action output $a$.}, respectively. $\mathcal{E}$ denotes the set of edges, where each edge $e$ is characterized by: its source and destination vertices $\nu_i$ and $\nu_j$, 
a feature vector $g_{ij}=[1/d_{ij},\chi_{ij}]^\top$, where $d_{ij}$ is a distance measure, introduced in \cite{klimke2022cooperative} and based on the Mahalanobis distance, while $\chi_{ij}$ is the bearing angle from vehicle $j$ to vehicle $i$. Finally, $r\in \mathcal{U}=\{\textit{same lane, crossing}\}$ specifies the edge type. An edge is established between two vehicles only if their trajectories require coordination for a safe intersection crossing. When vehicles are on different lanes but share a potential conflict point ahead, their corresponding nodes are linked bidirectionally with \textit{crossing} edges. Conversely, if two vehicles are traveling along the same path, a \textit{same lane} edge is introduced, directed from the leading vehicle to the one following behind.

\section{Proposed Approach}
\label{sec:method}

\subsection{Augmenting the state space}
Differently from what proposed in \cite{klimke2022enhanced, klimke2022cooperative}, and as already discussed in Section \ref{sec:intro}, the objective of this study is to tackle a \textit{mixed} autonomous intersection management scenario, where vehicles are heterogeneous and come from two different populations, i.e. \textit{petrol (p)} and \textit{electric (e)} vehicles, which is an increasingly common scenario in modern urban networks.

This shift of paradigm requires a modification of the standard graph-based state representation described in Section~\ref{subsec:graphaim}, in order to allow the RL agent to refine its strategies among heterogeneous vehicle populations, by understanding type-specific characteristics in the graph representation.
Taking inspiration from \cite{klimke2023automatic}, and adapting their intuitions to suit our specific objectives, we augment the feature vector of each vertex $\nu$ with a binary variable $k$, denoting its type, i.e. $k=0$ for a petrol vehicle and $k=1$ for an electric vehicle. Moreover, we also expand the edge type set, i.e. $\mathcal{U}=\mathcal{U}_{\text{rel}}\times \mathcal{U}_{\text{fuel}}$, where $\mathcal{U}_{\text{rel}}$ is equal to the original $\mathcal{U}$ presented in Section \ref{subsec:graphaim}, while $\mathcal{U}_{\text{fuel}}=\{pp, pe, ep, ee\}$ specifies the types of the vehicles originating the edge. 

\subsection{Multi-objective reward function design}
Given the inherent multi-objective nature of our problem, the design of a suitable reward function is a crucial yet very delicate step towards the development of an effective strategy. As in the majority of the RL-based solutions for AIM, we construct our final reward function by combining various terms, which encode the different objectives present in the autonomous intersection management environment. However, in this section we only present the different components, and we leave the discussion about how to effectively combine them to build our final multi-objective reinforcement learning strategy to Section \ref{subsec:omega}.

The three building blocks of our reward function are the following:
\begin{itemize}
	\item Efficiency term $R_{\text{eff}}$, to improve the throughput of the intersection by rewarding sustained speeds:
	\begin{align}
		R_{\text{eff}}=\frac{1}{|\mathcal{V}|}\sum_{\nu\in\mathcal{V}}\begin{cases}
			1.25\tfrac{v_\nu}{v_{\text{lim}}},\quad &\text{if }\tfrac{v_\nu}{v_{\text{lim}}}\leq 0.8; \\
			1.0, \quad &\text{if }0.8<\tfrac{v_\nu}{v_{\text{lim}}}\leq 1.0; \\
			6.0 - 5.0\tfrac{v_\nu}{v_{\text{lim}}}, \quad &\text{if }1.0\leq \tfrac{v_\nu}{v_{\text{lim}}};
		\end{cases}
	\end{align}
	where $v_{\text{lim}}$ denotes the speed limit.
	\item Environmental term $R_{\text{env}}$, designed to reduce CO$_2$ emissions produced by petrol vehicles:
	\begin{equation}
		R_{\text{env}}=-\frac{1}{|\mathcal{V}_p|}\sum_{\nu_p\in\mathcal{V}_p}\textsc{emission}(\nu_p),
	\end{equation}
	where $\mathcal{V}_p\subseteq\mathcal{V}$ is the set containing only the petrol vehicles, and \textsc{emission}$(\nu_p)$ is the amount of CO$_2$ emitted by vehicle $\nu_p$ in the considered time step.
	\item Safety term $R_{\text{saf}}$, to avoid collisions but at the same time the trivial solution of stopping the whole traffic:
	\begin{align}
		R_{\text{saf}}=
		\begin{cases}
			\begin{array}{rl}
				-10, & \quad \text{if a collision occurs;} \\
				-1,  & \quad \text{if all vehicles are standing still;} \\
				0,  & \quad \text{otherwise}.
			\end{array}
		\end{cases}
	\end{align}
\end{itemize}

\subsection{Learning multiple policies with a unique RL agent}
\label{subsec:omega}
Differently from previous studies, our objective goes beyond the development of a unique policy for autonomous intersection management. Instead, we endeavor to identify an entire spectrum of Pareto efficient policies, thereby facilitating adaptive policy adjustments in response to varying operational requirements, in terms of efficiency and environmental sustainability. Moreover, our approach obviates the necessity for retraining the algorithm from scratch whenever a new trade-off policy is required.

To achieve this goal, we first introduce a weighting factor $\omega\in[0,1]$, which allows us to define the overall parametric reward function in the following way:
\begin{equation}
	\label{eq:finalreward}
	R_\omega = \omega\cdot R_{\text{eff}} + (1 - \omega) \cdot R_{\text{env}} + R_{\text{saf}}.
\end{equation}
Instead of fixing a specific value for $\omega$ and learning a single policy, as it is usually done in literature, we let $\omega$ vary during the training phase. In particular, at each time step we sample a different $\omega$ from a uniform distribution, i.e. $\omega\sim\mathcal{U}[0,1]$. Finally, it is essential to refine also the state representation, by incorporating the parameter $\omega$. This augmentation ensures that the RL agent fully captures the changes in the reward function, and it also preserves the stationarity of the environment, a key assumption in~RL.

In conclusion, we can now train a unique RL algorithm that will learn a set of parametric policies, as $\omega$ varies between 0 and 1. Then, at execution time, the service providers will be able to choose the trade-off policy that best fits their needs, and they will also have the possibility of changing it over time, by simply modifying the parameter $\omega$ input to the policy. 
As a final remark, we note that the safety term $R_{\text{saf}}$ in equation \eqref{eq:finalreward} is included as an additive term that is not scaled or affected by $\omega$. This reflects the fact that safety is considered a non-tradable objective in the system. Including $R_{\text{saf}}$ outside the trade-off captures the idea that maintaining safe behavior is fundamental, regardless of how we tune the priorities between efficiency and environmental objectives.

\subsection{Fairness considerations}
\label{subsec:fair}

As already discussed in Section \ref{sec:intro}, introducing a trade-off between efficiency and environmental sustainability in a mixed AIM scenario inherently coincides with introducing unfairness of service among the different categories of road users. This happens because electric vehicles are not affected by the aforementioned trade-off, given that they do not emit any CO$_2$. For this reason, we must also take into consideration fairness criteria when deploying our strategy. In this work, we have decided to tackle the problem in a \textit{post-hoc} fashion: as we will see in Section \ref{sec:exp}, after building the frontier of Pareto efficient strategies, we will analyze their fairness score by measuring the average difference in travel time between an electric and a petrol vehicle, denoted by $\Delta_F$:
\begin{equation}
	\Delta_F = \cfrac{1}{|\mathcal{V}_p|} \sum_{\nu_p\in\mathcal{V}_p}T(\nu_p) - \cfrac{1}{|\mathcal{V}_e|} \sum_{\nu_e\in\mathcal{V}_e}T(\nu_e),
\end{equation}
where $\mathcal{V}_e\subseteq\mathcal{V}$ is the set containing only the electric vehicles and $T(\nu)$ is the travel time of vehicle $\nu$.

Finally, the guideline that we propose to effectively utilize our fairness-oriented MORL approach for AIM is the following: given an upper bound on the emission level that we can admit and a lower bound on the average speed that we want the vehicles to maintain, we select the fairest solution among the ones in the Pareto frontier. In this way we balance the trade-off between efficiency and environmental sustainability, and we also implement the solution with the lowest possible unfairness level between petrol and electric vehicles.

\section{Experiments}
\label{sec:exp}

% FIG 1
\begin{figure*}[h]
	\centering
	\begin{subfigure}[b]{0.48\textwidth}  % Adjust width as needed
		\centering
		\includegraphics[width=\textwidth]{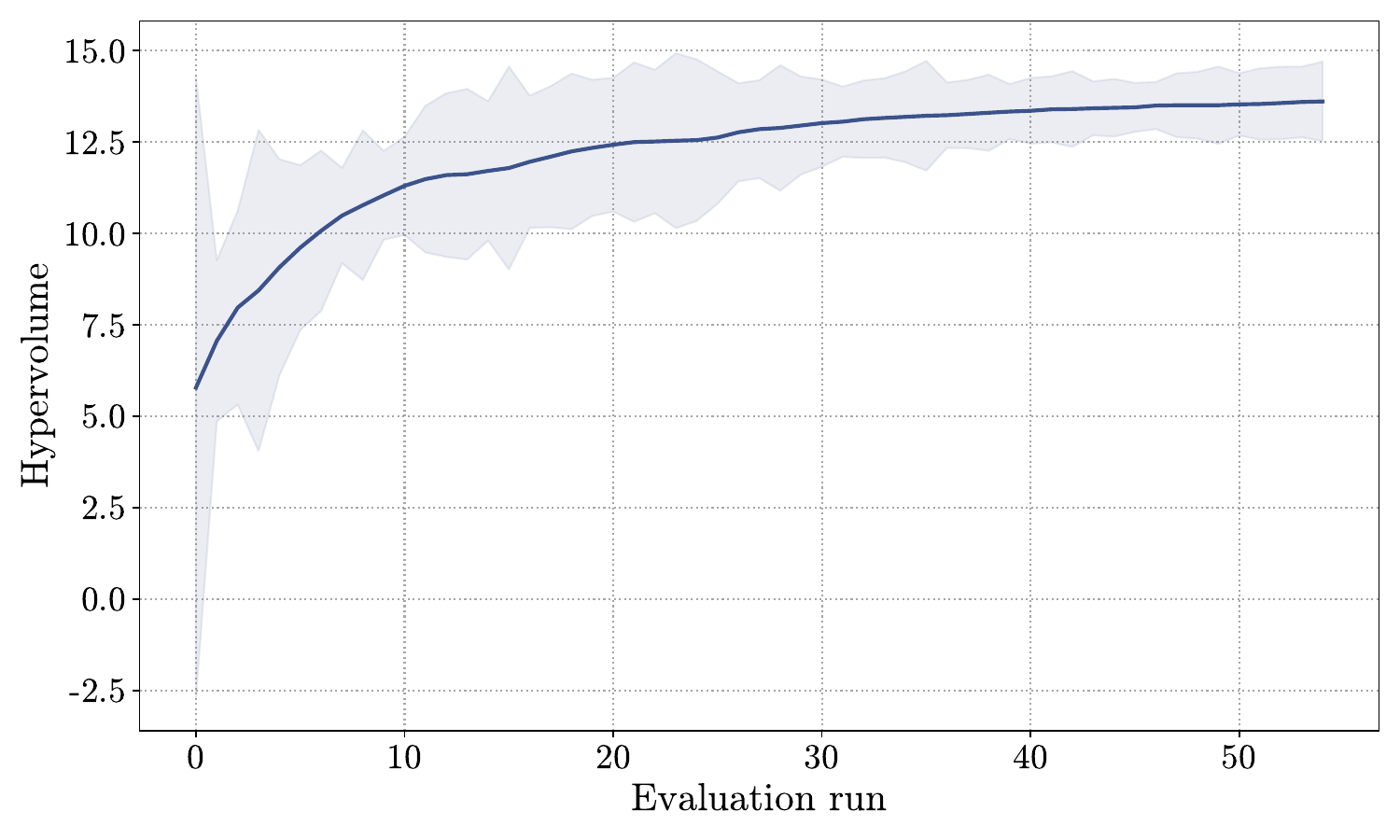}
		\caption{Hypervolume training curve}
		\label{fig:hv}
	\end{subfigure}
	\hfill
	\begin{subfigure}[b]{0.48\textwidth}  % Adjust width as needed
		\centering
		\includegraphics[width=\textwidth]{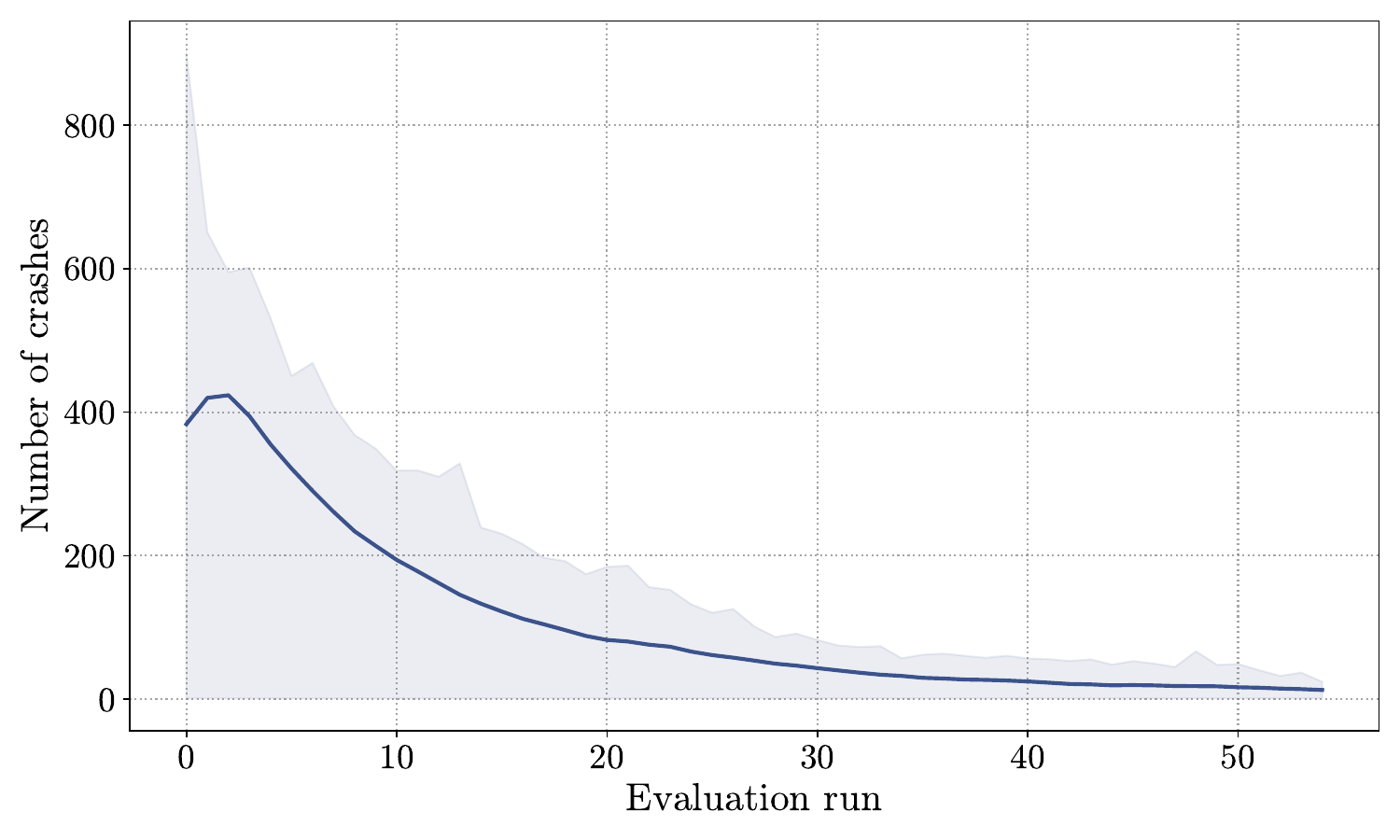}
		\caption{Number of crashes training curve}
		\label{fig:crashes}
	\end{subfigure}
	
	\caption{Training metrics of our approach. Figure \ref{fig:hv} shows the evolution of the hypervolume metric as a function of the evaluation runs performed during training. Similarly, Figure \ref{fig:crashes} displays the evolution of the number of crashes as training progresses. The shaded regions represent the $95\%$ confidence interval of the average evaluation over 10 random seeds.}
	\label{fig:training}
\end{figure*}

We begin this section by describing in detail the environment designed for our experiments and the graph neural network (GNN) architecture used to approximate $\pi_\phi$ and $Q_\theta$. Then, we will present and discuss the simulation results, to demonstrate the effectiveness of our approach\footnote{Code at: \url{https://github.com/mcederle99/FairMORL-AIM}}.

\subsection{Environment definition}
To effectively train and evaluate our algorithm, selecting an appropriate simulation environment is essential. For this project, we opted for the FLOW platform (\cite{wu2021flow}), specifically designed to support RL experiments focused on autonomous driving. FLOW integrates with SUMO (Simulation of Urban MObility, \cite{SUMO2018}), a popular microscopic traffic simulator available under an open-source license. In our setup, FLOW serves as the intermediary between SUMO and our RL framework by adhering to the standard Gymnasium APIs (\cite{towers_gymnasium_2023}), which are extensively utilized within the RL community.

The environment in which the agent acts is a four-way one-lane unsignalized intersection scenario, with three different turning intentions available to the vehicles coming from each of the four ways. Moreover, the simulation step has been fixed to 100ms, and the flow rate of vehicles coming across the intersection has been fixed to 1200 veh/hour, similarly to previous works (\cite{klimke2022enhanced, klimke2022cooperative}).

\begin{figure}[h]
	\begin{center}
		\begin{tikzpicture}[
			node distance=0.7cm and 0.4cm,
			every node/.style={font=\small},
			box/.style={rectangle, rounded corners, draw, fill=orange!30, minimum height=1cm, minimum width=0.8cm},
			block/.style={rectangle, rounded corners, draw, fill=orange!30, minimum height=2cm, minimum width=0.6cm},
			layer/.style={rectangle, rounded corners, draw, fill=blue!50, minimum height=2cm, minimum width=0.6cm},
			layerx/.style={rectangle, rounded corners, draw, fill=teal!50, minimum height=2cm, minimum width=0.6cm},
			arrow/.style={-Stealth, thick}
			]
			
			% Input Nodes
			\node(v_enc) {$h_\nu$};
			\node[below=of v_enc] (e_enc) {$h_e$};

			% Hidden Layers (aligned horizontally)
			\node[box, right=of v_enc] (conv1_v) {$v_{enc}$};
			\node[box, right=of e_enc] (conv1_e) {$e_{enc}$};
			\node[box, below=of conv1_e] (conv1_o) {$o_{enc}$};
			\node[left=of conv1_o] (o_enc) {$\omega$};
			
			% Next Layer
			\node[layer, right=of conv1_v, yshift=-0.6cm] (conv2) {$RGCN^+$};
			
			\node[layer, right=of conv2] (conv3) {$RGCN$};
			
			% Output Nodes
			\node[block, right=of conv3] (agg) {$FF$};
			\node[box, right=of agg] (dec) {$dec$};
			\node[right=of dec] (output) {$a$};
			
			% Labels
			%\node[above=0cm of v_enc] {$h^{(0)}$};
			%\node[above=0cm of e_enc] {$e^{(0)}$};
			%\node[above=0cm of conv1_v] {$h^{(1)}$};
			%\node[above=0cm of conv1_e] {$e^{(1)}$};
			\node[above=0cm of conv2] {$h^{(1)}$};
			\node[above=0cm of conv3] {$h^{(2)}$};
			\node[above=0cm of agg] {$h^{(3)}$};
			
			% Connections
			\draw[arrow] (v_enc) -- (conv1_v);
			\draw[arrow] (e_enc) -- (conv1_e);
			\draw[arrow] (o_enc) -- (conv1_o);
			\draw[arrow] (conv1_v) -- (conv2);
			\draw[arrow] (conv1_e) -- (conv2);
			\draw[arrow] (conv2) -- (conv3);
			\draw[arrow] (conv3) -- (agg);
			\draw[arrow] (agg) -- (dec);
			\draw[arrow] (conv1_o) -| (agg);
			\draw[arrow] (dec) -- (output);
			
		\end{tikzpicture}
		\caption{Architecture for the actor network. We refer to \cite{klimke2022enhanced} and \cite{schlichtkrull2018modeling} for the description of the graph layers $RGCN^+$ and $RGCN$, respectively. The critic network is analogous apart from the $dec$ layer which outputs a single value instead of the joint action for all the vehicles.} 
		\label{fig:gnn}
	\end{center}
\end{figure}
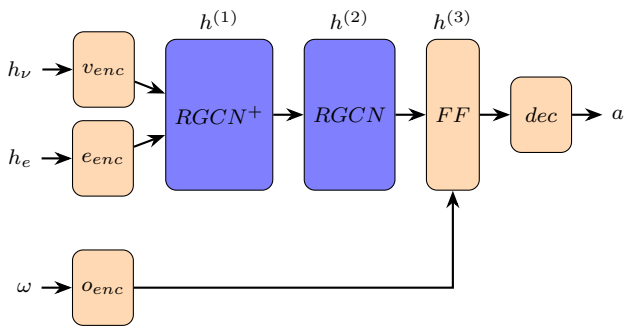

\subsection{Graph neural network architecture}
The architecture of the actor network is illustrated in Figure \ref{fig:gnn}. It follows a similar structure to the one proposed in \cite{klimke2022enhanced}, with the key distinction that it incorporates an additional state component, the trade-off parameter $\omega$. Consistent with the processing of vertex and edge features, we first apply a fully connected encoding layer to $\omega$ and then concatenate its output to the hidden feature vector $h^{(2)}$, produced by the graph layers. The resulting vector is subsequently passed through a fully connected layer before reaching the final decoding stage, which outputs a joint action $a$ for all vehicles in the environment.

\subsection{Simulation results}

% FIG 2
\begin{figure}[h]
	\begin{center}
		\includegraphics[width=8.4cm]{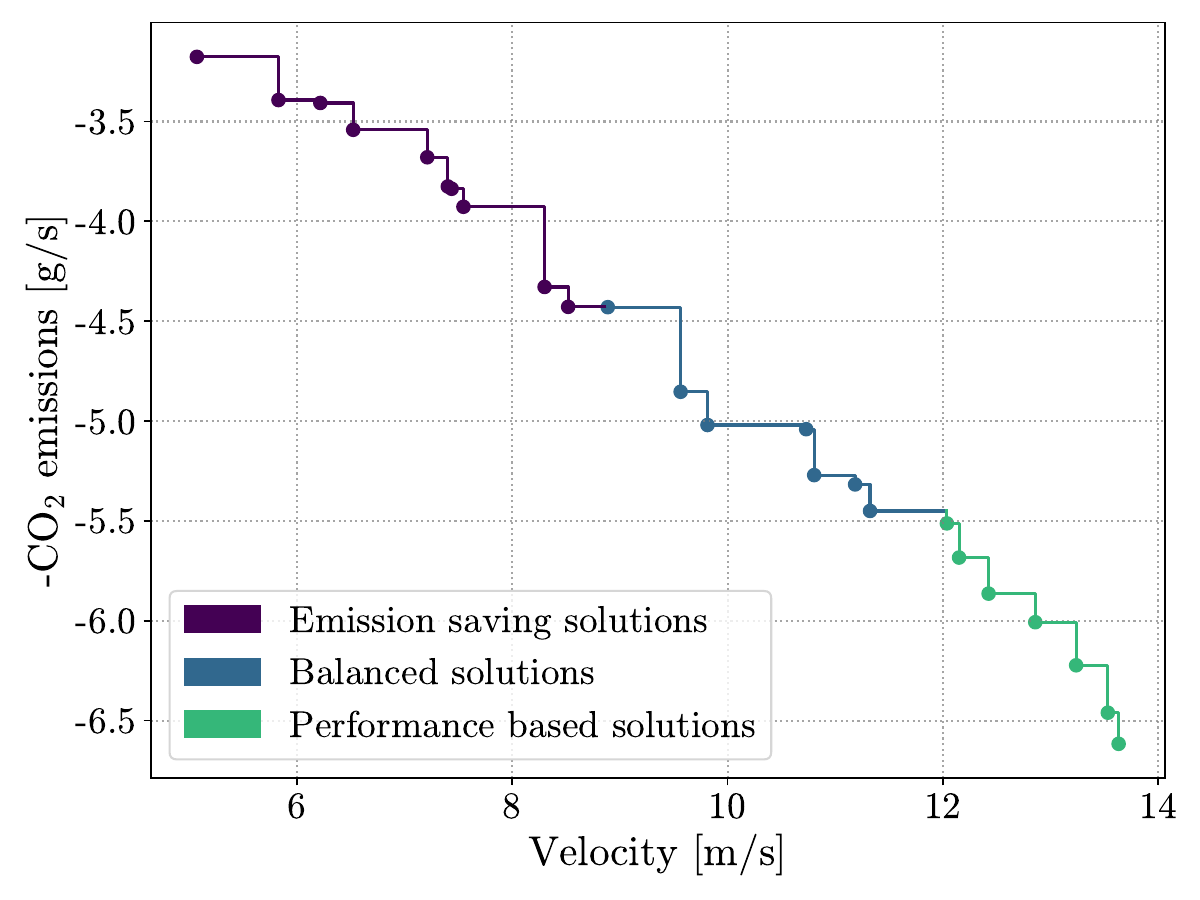}    % The printed column width is 8.4 cm.
		\caption{Pareto front for the considered MORL problem. The average efficiency maximization and emissions minimization objectives are represented on the x and y axes, respectively. Each point on the front corresponds to a different value of $\omega$.} 
		\label{fig:pareto}
	\end{center}
\end{figure}

% FIG 3
\begin{figure*}[t]
	\centering
	\begin{subfigure}[b]{0.46\textwidth}  % Adjust width as needed
		\centering
		\includegraphics[width=\textwidth]{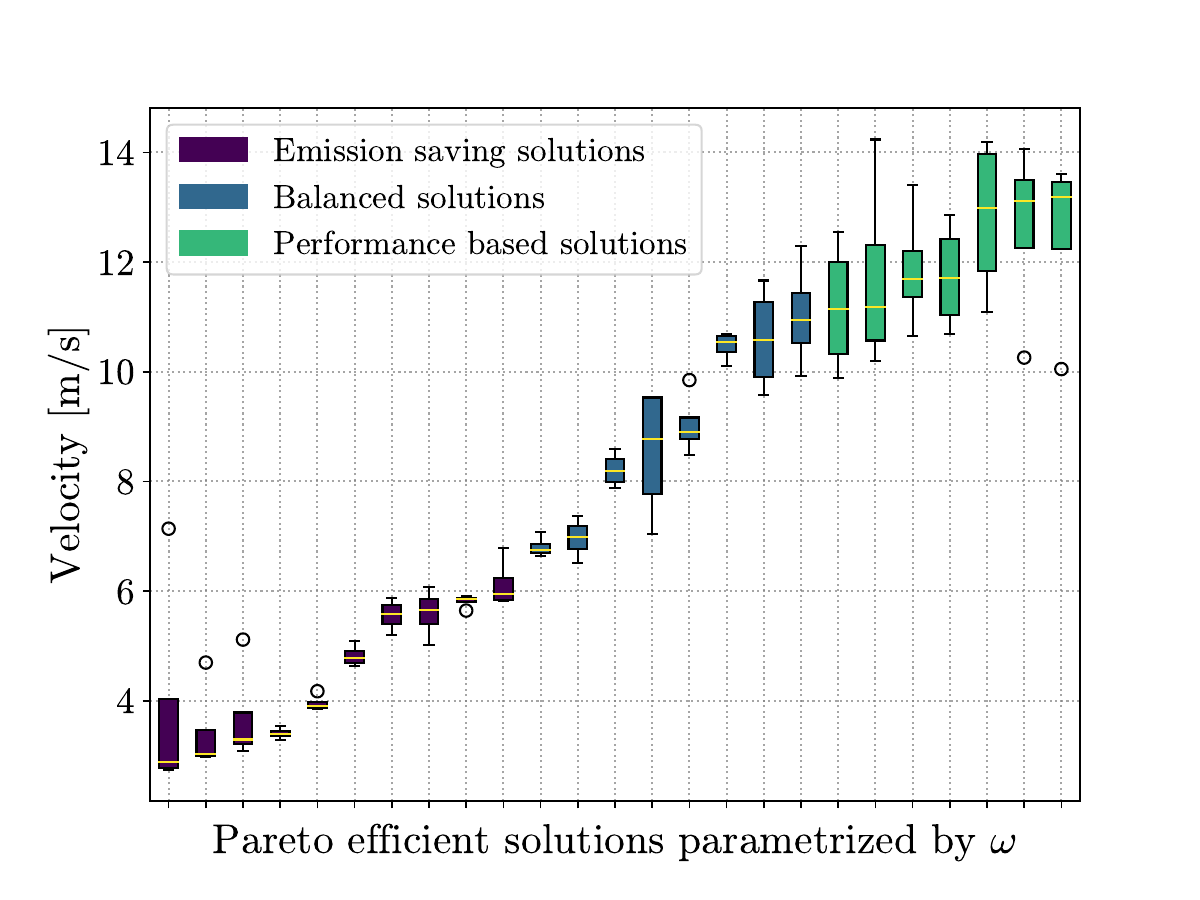}
		\caption{Velocities distribution at test time}
		\label{fig:boxvel}
	\end{subfigure}
	\hfill
	\begin{subfigure}[b]{0.46\textwidth}  % Adjust width as needed
		\centering
		\includegraphics[width=\textwidth]{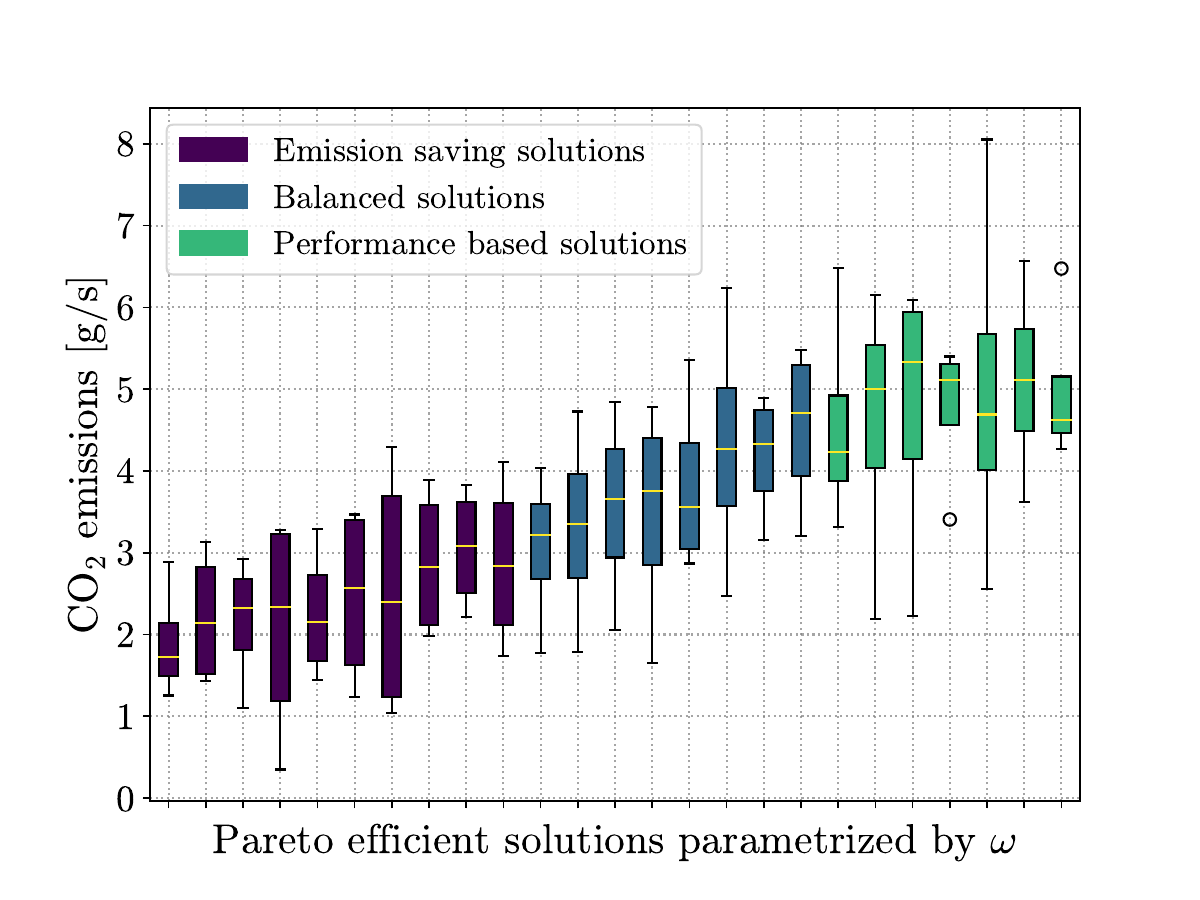}
		\caption{Emissions distribution at test time}
		\label{fig:boxem}
	\end{subfigure}
	
	\caption{Velocities and emissions distributions per time step during evaluation. Each point on the x axis corresponds to a Pareto efficient solution, going from left to right when looking at Figure \ref{fig:pareto}.}
	\label{fig:boxplots}
\end{figure*}

% FIG 4
\begin{figure}[h]
	\begin{center}
		\includegraphics[width=7.1cm]{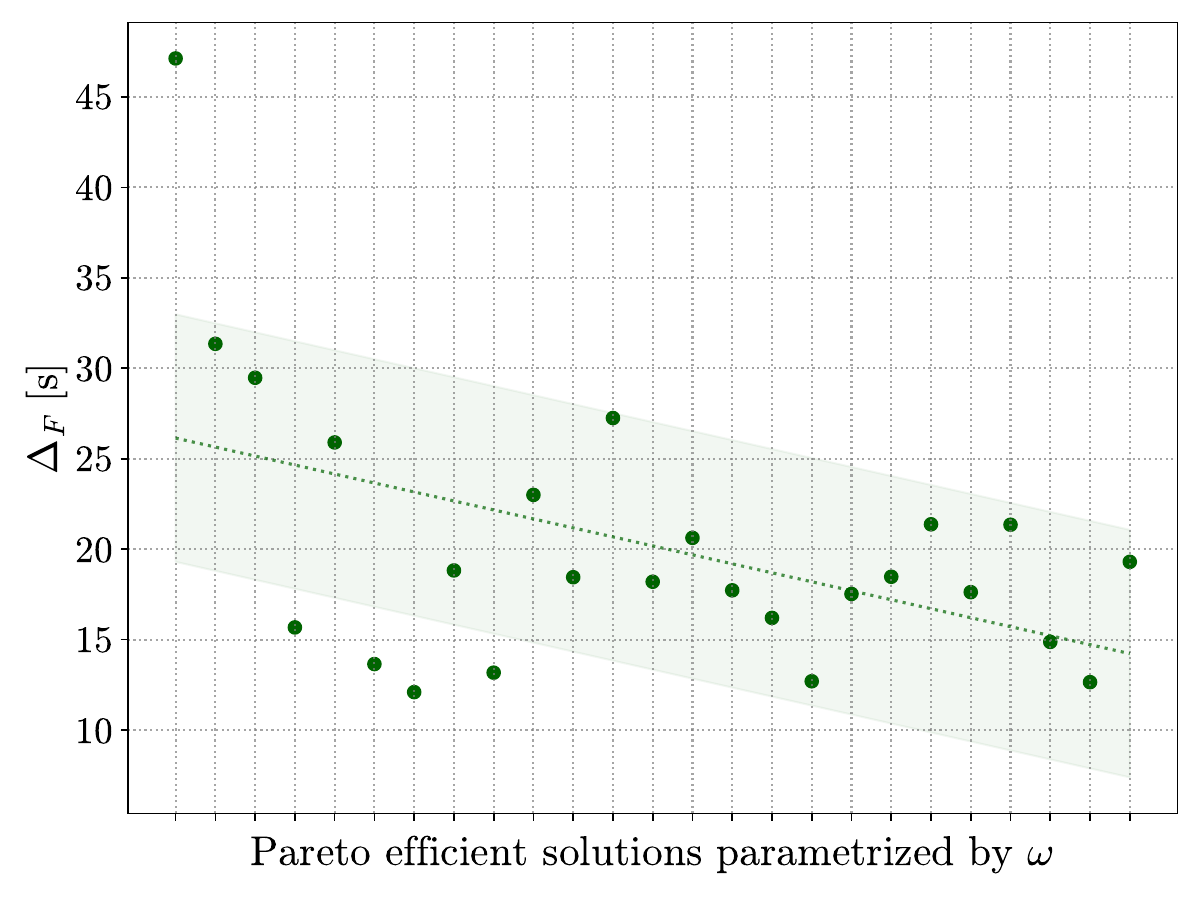}    % The printed column width is 8.4 cm.
		\caption{Average difference in travel time between electric and petrol vehicles. Each point on the x axis corresponds to a Pareto efficient solution, going from left to right when looking at Figure \ref{fig:pareto}. The dotted line and the shaded areas represent the linear interpolation of the points and their 95\% confidence interval, respectively.} 
		\label{fig:fairness}
	\end{center}
\end{figure}

To evaluate the learning performances of our algorithm, we trained it for $300k$ time steps using 10 different random seeds, and every $5k$ steps we paused training and run an evaluation phase\footnote{An evaluation phase consists in simulating $10k$ steps where the RL agent exploits its policy with no exploration. During this period, $\omega$ is periodically modified, increasing from 0 to 1.}, where we recorded two performance metrics, namely the hypervolume and the number of crashes. While the latter is trivial to explain, the former requires some further considerations. Indeed, unlike standard RL, MORL seeks to learn a set of policies that form a Pareto front rather than a single optimal policy. Consequently, traditional metrics like episodic return over training time are not suitable in our scenario. Instead, we use the hypervolume metric, widely used in the MORL field, which measures the volume of the region enclosed by the approximated Pareto front and a reference point in the objective space, providing a comprehensive assessment of the trade-offs between conflicting objectives (\cite{felten2024multi}). As clear from the plots in Figure \ref{fig:training}, both metrics have a consistent trend, showcasing the learning capabilities of our approach. As a further remark, and as expected from the literature, we recall that the RL agent  cannot completely eliminate collisions. However, the rare failure cases are not an issue in practice, because the cooperative maneuver will only be advertised to  the connected vehicles if it fulfills sanity checks like being collision-free; otherwise the connected vehicles would simply resort to local planning~(\cite{klimke2022enhanced, klimke2022cooperative}).

Moving further%on from the training metrics needed to assess convergence
, we now present the most insightful results from our study. Figure \ref{fig:pareto} shows the Pareto front obtained by performing an evaluation phase using the trained multi-objective RL agent. Each point on the front corresponds to a Pareto efficient policy, as $\omega$ varies between 0 and 1. As expected, the more $\omega$ increases, the more the corresponding policy is biased towards efficiency-oriented solutions. Conversely, when $\omega$ is low, the obtained policies are more focused on reducing the emissions. 
In addition to that, we have used the K-Means algorithm (\cite{jin2010k}) to cluster the Pareto efficient policies into 3 groups—emission saving, balanced, and performance-based solutions. Clustering techniques like these, often combined with probabilistic models such as Gaussian Mixture Models, have also shown strong potential for interpretability and behavior modeling in sensor networks (\cite{varotto2022visual}). We believe this step helps intersection managers choose the most suited policy for each specific situation.

%In addition to that, we have used the K-Means algorithm~(\cite{jin2010k}) to cluster the Pareto efficient policies into 3 groups, i.e. \textit{emission saving}, \textit{balanced} and \textit{performance based} solutions. We believe that this step is useful to enhance the interpretability of the model and help the intersection manager in choosing the most suited policy for each specific situation.

Finally, to offer deeper insights into the robustness of the obtained Pareto efficient solutions, we reported in Figure \ref{fig:boxplots} the boxplots of both the velocities and emissions per time step for all the policies in the front. As expected, both of them are increasing with the value of $\omega$, consistently with the above discussion.

To conclude our analysis, we have reported in Figure \ref{fig:fairness} the \textit{post-hoc} fairness metric introduced in Section \ref{subsec:fair}, i.e. $\Delta_F$. As visible from the plot, $\Delta_F$ exhibits a decreasing trend as $\omega$ increases; this is expected, since the less we care about the environmental emissions, the more petrol vehicles are treated similarly to electric vehicles. However, as we can see, due to the high complexity of the environment, the fairness metric is not monotonically decreasing as $\omega$ increases. This behavior is notable and worth studying when choosing the policy to deploy in the system, as we envision to explore in future studies. 

As already discussed in Section \ref{subsec:fair}, the guideline of choosing the fairest solution among the ones that respect some desired constraints on both efficiency and emissions seems to be the most suitable choice in this scenario.

\section{Conclusions}
\label{sec:conc}

We believe that this study represents a step forward in the autonomous intersection management field, thanks to the introduction of a novel multi-objective reinforcement learning framework. Our method was evaluated in a complex simulated scenario, where it demonstrated its ability of learning a diverse collection of Pareto optimal policies, which focus on finding the balance between the optimization of both efficiency and environmental sustainability.

In addition to that, this research introduces a crucial yet often undervalued fairness of service criterion, which serves as an essential tool for identifying the most equitable policy among the ones composing the Pareto front. This criterion ensures that the final deployed policy maintains the highest possible level of fairness across all users of the intersection, thereby enhancing the overall robustness and applicability of our approach in realistic scenarios.

Summarizing, our contributions underscore the potential of multi-objective reinforcement learning to transform intersection management, by not only improving operational efficiency but also fostering sustainable practices and equitable service provision, contributing positively to smarter and greener urban mobility solutions.
In future works we envision to extend this framework to more complex urban traffic environments and also to directly include the fairness objective into the formulation of the problem.

%\begin{ack}
%Place acknowledgments here.
%\end{ack}

%\bibliography{ifacconf}
\bibliography{REFERENCE_LIST_morl_for_aim}

%\appendix
%\section{A summary of Latin grammar}
%\section{Some Latin vocabulary}

\end{document}